\documentclass[%
 aip,
 amsmath,amssymb,
 reprint,%
]{revtex4-1}

\usepackage{graphicx}% Include figure files
\usepackage{dcolumn}% Align table columns on decimal point
\usepackage{bm}% bold math
\usepackage[breaklinks=true]{hyperref}
\hypersetup{bookmarksnumbered, pdfpagemode=UseOutlines, 
pdfdisplaydoctitle, 
colorlinks=true, citecolor=blue, filecolor=blue, linkcolor=blue, urlcolor=blue}
\usepackage[utf8]{inputenc}
\usepackage[T1]{fontenc}
\usepackage{mathptmx}
\usepackage{etoolbox}
\usepackage{comment}

\newcommand*{\affilmpsd}{Max Planck Institute for the Structure and Dynamics of Matter, Center for Free Electron Laser Science (CFEL), Luruper Chaussee 149, 22761 Hamburg, Germany}

\newcommand*{\affilcui}{The Hamburg Centre for Ultrafast Imaging, Luruper Chaussee 149, 22761 Hamburg, Germany}

\newcommand*{\affilaachen}{Institut f\"ur Theorie der Statistischen Physik, RWTH Aachen University and JARA-Fundamentals of Future Information Technology, 52056 Aachen, Germany}

\makeatletter
\def\@email#1#2{%
 \endgroup
 \patchcmd{\titleblock@produce}
  {\frontmatter@RRAPformat}
  {\frontmatter@RRAPformat{\produce@RRAP{*#1\href{mailto:#2}{#2}}}\frontmatter@RRAPformat}
  {}{}
}%
\makeatother
\begin{document}

\preprint{AIP/123-QED}

\title[Cavity Quantum Materials]{Cavity Quantum Materials}
% Force line breaks with \\
\author{F. Schlawin}
%\email{frank.schlawin@mpsd.mpg.de}
\affiliation{\affilmpsd}
\affiliation{\affilcui}%
 %\altaffiliation[Also at ]{Physics Department, XYZ University.}%Lines break automatically or can be forced with \\
\author{D. M. Kennes}%
%\email{dante.kennes@rwth-aachen.de}
\affiliation{\affilaachen}%
\affiliation{\affilmpsd}%

\author{M. A. Sentef}
 %\homepage{https://www.Second.institution.edu/~Charlie.Author.}
 \email{michael.sentef@mpsd.mpg.de}
\affiliation{%
\affilmpsd
}%

\date{\today}% It is always \today, today,
             %  but any date may be explicitly specified

\begin{abstract}
The emergent field of cavity quantum materials bridges collective many-body phenomena in solid-state platforms with strong light-matter coupling in cavity quantum electrodynamics (cavity QED). This brief review provides an overview of the state of the art of cavity platforms and highlights recent theoretical proposals and first experimental demonstrations of cavity control of collective phenomena in quantum materials. This encompasses light-matter coupling between electrons and cavity modes, cavity superconductivity, cavity phononics and ferroelectricity, correlated systems in a cavity, light-magnon coupling, cavity topology and the quantum Hall effect, as well as superradiance. An outlook of potential future developments is given.
\end{abstract}

\maketitle
\tableofcontents

\section{Introduction}
Quantum materials are complex quantum many-body systems consisting of multiple atomic species. Their low-energy physics typically features a complex interplay of charge, spin, orbital, and lattice degrees of freedom, giving rise to intricate phase diagrams with correlated and/or topological ground states that can be controlled by external stimuli. The control of quantum materials through nonthermal pathways with short and strong laser pulses has been reviewed recently \cite{de_la_torre_colloquium_2021,disa_engineering_2021}. 

A different strategy towards modifying emergent properties of quantum materials is opened when one considers the replacement of a classical laser field by quantum-mechanical photon modes in a cavity. Strong light-matter couplings were demonstrated in molecules \cite{chikkaraddy_single-molecule_2016,santhosh_vacuum_2016}, evidenced by large Rabi splittings, and have led to first demonstrations and proposals of cavity-induced changes of chemical reactions, sometimes coined ``polaritonic chemistry'' \cite{feist_polaritonic_2018,ruggenthaler_quantum-electrodynamical_2018,ribeiro_polariton_2018,flick_strong_2018,reitz_cooperative_2021}. Moreover the enhancement of charge transport in organic molecules in cavities \cite{orgiu_conductivity_2015} was reported and spurred subsequent theoretical investigations into cavity-induced changes in transport \cite{schachenmayer_cavity-enhanced_2015,hagenmuller_cavity-enhanced_2017,hagenmuller_cavity-assisted_2018} and optical properties \cite{rokaj_free_2021,eckhardt_quantum_2021,amelio_optical_2021}. This led to the suggestion that condensed-matter platforms and their embedding in cavities should be further explored with the goal to modify their emergent properties \cite{ebbesen_hybrid_2016, Garcia-Vidal2021}.

While light-matter coupling in the classical case is weak, given by the bare fine structure constant in vacuum, many photons created by a laser occupy a macroscopic coherent state that can change a material's properties. On the contrary, the key idea behind the engineering of materials' properties with quantized photon modes is that the effective optical mode volume can be reduced by constructing a suitable surrounding (cavity), and the light-matter coupling can thus be increased compared to its bare value (ultrastrong or even deep strong coupling, see Sec.~\ref{sec:cavity_QED}). Due to the stronger coupling the number of photons required to significantly change the physics of the light-matter system can potentially be reduced, perhaps even to the point where no photons are required (dark cavity) and the pure quantum fluctuations of light suffice to achieve certain effects.  

The purpose of this brief review is to provide an overview of recent research activities that bridge the fields of quantum materials and cavity quantum electrodynamics, which we coin ``cavity quantum materials''. Importantly, this article is not aiming at providing an extensive review of research into strong cavity light-matter coupling in all possible platforms. 
In particular, we note that there is an entire field of exciton polaritonics and exciton-polariton condensates, which is complementary to the focus of this brief review in that the coupling to light happens via quasi-bosonic excitations of the electron system, namely excitons, relevant mostly in semiconductors. We refer to excellent reviews of the rich phenomenology in this field for further reading \cite{deng_exciton-polariton_2010,byrnes_excitonpolariton_2014}. Moreover, ultracold quantum gases constitute another important platform in which collective behavior of cavity-coupled many-body systems can be studied; here we refer to a recent comprehensive review, Ref.~\onlinecite{Mivehvar2021}. 

This brief review is structured as follows. First, we give an overview of the state of cavity quantum electrodynamics (cavity QED) insofar as it concerns cavity control of materials in Sec.~\ref{sec:cavity_QED}. Then we discuss the cavity control of quantum materials in Sec.~\ref{sec:quantum_materials}. We focus in particular on the different light-matter coupling opportunities in Sec.~\ref{subsec:LMC}, and the cavity control of collective phenomena in Sec.~\ref{subsec:collective}%, and the complementary field of cavity excitonics and polaritonics in Sec.~\ref{subsec:excitonics}
. We provide an Outlook of promising directions and relevant future work in Sec.~\ref{sec:outlook}.

\section{Cavity QED and experimental state of the art}
\label{sec:cavity_QED}
%FRANK
%Definitions of strong, ultrastrong, deepstrong
%Types of cavities

\begin{figure*}[t]
\centering
\includegraphics[width=0.65\textwidth]{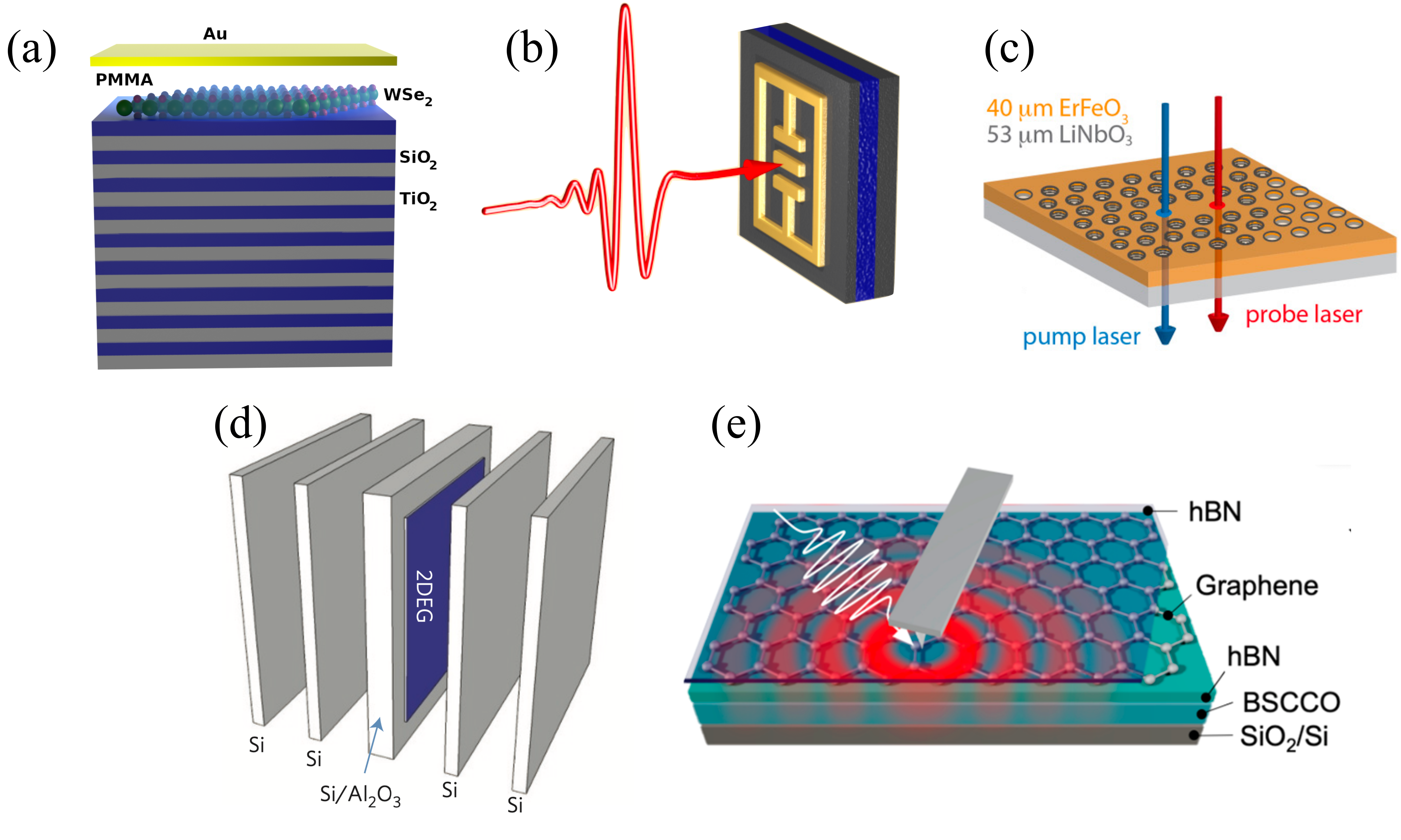}
\caption{
\textbf{Platforms in condensed matter cavity QED} 
(a) Microcavity setup for the coupling of interband transitions. In this hybrid example, the top cavity mirror consists of a gold sheet. The dielectric PMMA filling material fills the cavity volume. The sample is given by a monolayer semiconductor, here WSe$_2$. The bottom mirror of the resonator is formed by a distributed Bragg reflector (DBR). Such a setup was employed, e.g., in Ref.~\onlinecite{Lundt2016}. 
(b) A nanoplasmonic split ring cavity, where the resonator is formed by the gold structure. The field is mostly confined between the two horizontal bars, and can couple evanescently to an underlying two-dimensional material, here sketched as the blue layer. Figure reprinted with permission from Nano Letters 17, 6340–6344 (2017), %\url{https://doi.org/10.1021/acs.nanolett.7b03103},
Ref.~\onlinecite{Bayer2017}. Copyright 2017 American Chemical Society. 
(c) and (d) Two realizations of terahertz photonic crystal cavities. In panel (c), the cavity is formed by defects in a two-dimensional array of holes in the dielectric material. Figures (c) and (d) reprinted from Journal of Applied Physics 125, 213103 (2019), Ref.~\onlinecite{Nelson2019}, with the permission of AIP publishing. In (d), Silicon wafers act as Bragg reflectors to contain the cavity field. Figure reprinted with permission from Nature Physics 12, 1005-1011 (2016),  Ref.~\onlinecite{Kono2016}. Copyright 2016 Springer Nature. 
(e) Proposal of a graphene/hBN surface plasmon cavity coupling to quantum material (BSCCO) below. 
Reprinted with permission from Nano Letters 21, 308–316
(2021), Ref.~\onlinecite{Berkowitz2021}. Copyright 2021 American Chemical Society.
}
\label{fig.QED-setup}
\end{figure*}

Cavity QED traditionally concerns itself with the interaction of a single or multiple emitters with the well-defined field mode of a cavity. If only a single dipolar transition of the emitters couples strongly to the cavity with frequency $\omega$, we can describe it in terms of a simple two-level model and quantify the light-matter interaction by a single coupling strength $g$. 
In order to assess the emerging physics, we have to compare this coupling strength with the loss rates in the system, i.e., with the cavity loss $\gamma_{cav}$ and the material loss $\gamma_{mat}$. The cavity quality factor is defined as
%\begin{equation} 
 %   Q = \frac{\omega}{\gamma_{cav}}.
%\end{equation}
$Q = \omega / \gamma_{cav}$.

In the \textbf{weak coupling} regime, where the loss rates dominate over the coherent coupling, $g \ll \gamma_{cav}, \gamma_{mat}$,
the dynamics are essentially incoherent and each emitter will interact independently with the cavity. Nevertheless, the presence of the cavity manifests itself in the change of the excited state lifetimes of the emitters as described by the Purcell effect \cite{Purcell-effect}. 

When the coherent coupling becomes comparable to the loss rates, $g \gtrsim \gamma_{cav}, \gamma_{mat}$, we enter into the \textbf{strong coupling} regime, where the recurrent exchange of excitation quanta between cavity mode and emitters leads to the formation of hybridised states which inherit properties of both constituent parts \cite{FornDiaz2019,frisk_kockum_ultrastrong_2019}. 
In this regime, the different emitters start to feel each other's presence, forming collective states called polaritons \cite{ballarini_polaritonics_2019}. The presence and stability of such collective states is usually quantified by the
cooperativity $C = 4 g^2 / (\gamma_{cav} \gamma_{mat})$  \cite{FornDiaz2019,reitz_cooperative_2021}.
%\begin{equation}
 %   C = \frac{4 g^2}{\gamma_{cav} \gamma_{mat}}.
%\end{equation}
A high cooperativity $C\gg 1$ signals a regime where emitters behave as a single collective state.

When the coupling becomes even stronger such that it is comparable to the bare cavity frequency, $g \lesssim \omega$, we enter the \textbf{ultrastrong coupling} regime. In this regime, counter-rotating, energy-nonconserving terms in the light-matter interaction become relevant and the ground state of the coupled cavity-emitter system is affected. Recent theory work indicates that this can even happen in the presence of strong losses~\cite{DeLiberato2017}.

Finally, recent experiments have reached an even more extreme regime, dubbed \textbf{deep strong coupling}, where the coherent coupling becomes the dominant energy scale in the system \cite{Bayer2017, FornDiaz2019}. In particular, it is larger than the bare cavity frequency, $g > \omega$. 

With current technologies, there are two approaches to reach a strong light-matter coupling regime: 1) In the optical regime, where the coupling is always a fraction of the bare cavity frequency, $g\ll \omega$, it can only be reached in high-finesse cavities with large quality factors \cite{Walther2006, Aspelmeyer2014} in order to beat the loss rate. 2) More recently, with the advent of nanoplasmonic cavities in the terahertz or microwave regime, another route to reach a strong coupling regime has become available: it leverages the strong compression of plasmonic nearfields far below the farfield diffraction limit, $V_{mode} \ll (\lambda/2)^3$, to create very large interaction strengths. These can reach strong coupling (or even deep strong coupling \cite{Bayer2017}) in the presence of "noisy" samples with large losses or low quality factors. It is this second route which is most appealing for the manipulation of quantum materials, and in the following we will review the different platforms that are existing today.

%** What about: superconducting microwave cavities -> deep strong coupling **

\paragraph{Microcavities}

Microcavity structures are multimode cavities which are formed by metallic or dielectric mirrors~\cite{Kavokin2017}. We depict a hybrid example in Fig.~\ref{fig.QED-setup}(a). An in-depth introduction to the coupling of high-quality microcavities to monolayer semiconductors can be found in Ref.~\onlinecite{Schneider2018}.
Metallic microcavities are intrinsically lossy, thus restricting the achievable quality factor, but their strong confinement of the cavity light field produces large Rabi splittings. Conversely, distributed Bragg reflectors (DBRs) allow for higher quality factors, while limiting the confinement.

\paragraph{Split-ring resonator cavities}

An example of a nanoplasmonic split-ring resonator is depicted in Fig.~\ref{fig.QED-setup}b). 
These gold structures can be coupled to two-dimensional electron gases through evanescent waves \cite{scalari_ultrastrong_2012}, and are capable of achieving the largest cavity mode volume compression reported to date, $V \sim 10^{-5} \lambda^3$. They are consequently the only experimental platform, where coupling strengths exceeding the bare cavity resonance were reported~\cite{Bayer2017, Halbhuber2020}, thus reaching the deep strong coupling regime. Their metallic nature restricts their quality factor to below $Q\sim 10$.
%, even though superconducting analogues are
A detailed discussion of their properties is found in Ref.~\onlinecite{Maissen2014}. 
These platforms were recently used to induce strong coupling of collective excitations in cuprate superconductors with light \cite{schalch_strong_2019,keiser_structurally_2021}.

\paragraph{Array defect cavities}

Fig.~\ref{fig.QED-setup}(c) shows an example of a terahertz defect cavity. It is formed by defects in a two-dimensional array of holes, which are drilled inside a dielectric material. Here the cavity field can couple to phononic or magnonic excitation inside this material \cite{Nelson2019}.
These cavities combine substantial mode compression compared to free space cavities, $V \sim 10^{-3} \lambda^3$ with larger quality factors than split-ring cavities, for which $Q \lesssim 10^3$.

\paragraph{Semiconductor heterostructure cavities}

Fig.~\ref{fig.QED-setup}(d) shows another example of a photonic crystal cavity. It is composed of silicon wafers which form DBRs, and a central block acting as a defect where large field enhancements can be achieved. A two-dimensional electron gas, as indicated in Fig.~\ref{fig.QED-setup}(d), which is placed near this central block can experience large light-matter coupling at large cooperativities~\cite{Kono2016}, such that energy non-conserving terms in the interaction become relevant~\cite{Kono2018}.

\paragraph{Plasmonic/polaritonic layer cavities}

Strong light-matter coupling can be achieved by placing a sample in the vicinity of a metallic, graphene or hexagonal boron nitride (hBN) layer, where strongly confined surface plasmon polaritons can hybridise with the sample's excitations~\cite{Baumberg2019}. A theoretical proposal for such a structure is shown in Fig.~\ref{fig.QED-setup}(e). 
%Extreme nanophotonics from ultrathin metallic gaps (review) \cite{Baumberg2019}
A recent experiment, Ref.~\onlinecite{Epstein2020}, reported the largest mode volume compression to date, $V \sim 5 * 10^{-10} \lambda_0^3$ (where $\lambda_0$ is the free space resonance wavelength) in the mid-IR and Thz region.
This platform also enables strong coupling with propagating phonon polaritons \cite{Bylinkin2021}, i.e. coupling at large momenta.

\paragraph{Terahertz Fabry-P\'erot cavities}

Cryogenic cavities in the terahertz regime were recently developed\cite{Jarc2021}. While they do not feature the strong field compression of nearfield cavities, their unique strength lies in the continuous tunability of the cavity resonances, which will enable their coupling to a great variety of collective modes in correlated materials.

%\paragraph{Plasmonic nearfield cavities}

\section{Cavity control of quantum materials}
\label{sec:quantum_materials}

\subsection{Cavity light-matter coupling}
\label{subsec:LMC}
Fundamentally, charged matter couples to light (or any form of electromagnetic field) via the minimal coupling, i.e., $ \bm{p} \to  \bm{p}-q \bm{A} $, where $\bm{p}$ is the momentum of the particle, $q$ is its charge and $\bm{A}$ is the vector potential of the electromagnetic field. Here, we have set the scalar potential to zero, describing the light field by $\bm{A}$ alone. Freely moving electrons with charge $q=-e$ thus experience a coupling described via the kinetic energy
\begin{align}
     \frac{\bm{p}^2}{2m}\to \frac{1}{2m} (\bm{p}+e\bm{A})^2&=\frac{1}{2m}\left(\bm{p}^2+e\bm{p}\bm{A}+e\bm{A}\bm{p}+e^2 \bm{A}^2\right)\notag\\
     &=\frac{1}{2m}\left(\bm{p}^2+2e\bm{A}\bm{p}+e^2\bm{A}^2\right),
\end{align}
where for the last equality we have chosen the Coulomb gauge $\nabla \bm{A}=0$ for convenience, yielding $\bm{p}\bm{A}=\bm{A}\bm{p}$. In general both terms, $e\bm{A}\bm{p}$ and  $e^2\bm{A}^2$, must be kept, although for small $\bm{A}$ (or $e$) the latter is often neglected over the former. Such a situation is not applicable in the QED strong light-matter coupling context we are considering here. 

The example of freely moving charges discussed above, though instructive, is not very relevant to the case of materials strongly coupled to light in a cavity. Firstly, the ions of the crystal (assumed within a Born approximation to be static and classical) define a periodic lattice potential requiring the extension of Bloch's theorem in the presence of a cavity mode. Secondly, fluctuations in the charged ions' position (even if the Born approximation holds to high precision) can couple to the light field as well, yielding direct phonon-photon coupling. Thirdly, electronic excitation via the light field can excite a myriad of collective phenomena (being of collective electronic, vibrational, or other nature); a subsidiary effect which requires a detailed study of the microscopic interplay of different degrees of freedom in a solid. It can be described using the minimal coupling approach upon integrating out the fermionic degrees of freedom \cite{curtis_cavity_2019,allocca_cavity_2019,raines_cavity_2020}.

To address the first complication of electrons moving in a periodic potential subject to a light field, we need to extend the standard Bloch theorem to include the, in general time- and space-dependent, vector potential $\bm{A}$ \cite{rokaj_quantum_2019}. Furthermore, in the case of strong light-matter coupling in a cavity the quantization of the light field might be of relevance and the classical field $\bm{A}$ should be replaced by a quantized one. In a pragmatic approach (valid in the regime of not too strongly varying vector potentials and not too strong light matter-interaction) one might start with a tight-binding model for the material and add the effects of the light field by Peierls substitution where hopping amplitudes between Wannier states $\nu$ and $\nu'$ centered at $R$ and $R'$ are dressed by a phase factor (written here for spatially homogeneous field, i.e., in dipole and single-mode approximation),
\begin{equation}
    t_{\nu,\nu'}(R-R')\to t_{\nu,\nu'}(R-R') e^{-i\frac{e}{\hbar c}\bm{A}(R-R')},
\label{eq:peir}\end{equation}
and the field $\bm{A}$ is quantized after this substitution has been performed. In tight-binding models, further light-matter coupling terms can arise compared to the case when free particles are minimally coupled to a gauge field. One of these intriguing additional light-matter couplings was recently linked to the quantum geometry of the electronic wavefunctions in the presence of non-trivial quantum geometry \cite{PhysRevB.104.064306,Li2021,Ahn2021}. These couplings go entirely beyond a classical paradigm, where only the electronic band slope (velocity) and curvature (inverse effective mass) determine the paramagnetic and diamagnetic light-matter couplings, respectively. In certain cases, the quantum-geometric terms can dominate the physics, in particular in flat-band systems \cite{PhysRevB.104.064306} where classical light-matter couplings vanish. Intriguingly, in these flat-band systems the vanishing electronic kinetic energy can thus lead to comparatively large quantum-geometric light-matter coupling effects even when the light-matter coupling strength is not large per se.  

%In the ultrastrong coupling regime the above questions, e.g., how to set up tight-binding models, have attracted a lot of research interest lately \cite{Zueco2021b,rokaj_free_2021,li_electromagnetic_2020,dmytruk_gauge_2021,Ashida2021}. In this limit, solving the full problem of a light-matter coupled particle in a periodic potential proves more difficult \cite{Ashida2021}, and it is therefore not necessarily straightforward that the simple Peierls substitution of Eq.~\eqref{eq:peir} and subsequent quantization of the vector potential is the correct procedure to include the light-matter coupling\cite{Ashida2021, ashida_quantum_2020}. One successful strategy in the limit where light-matter interaction becomes the dominant energy scale was put forward in Ref.~\onlinecite{Ashida2021} and relies on a unitary transformation that achieves asymptotic decoupling of light and matter degrees of freedom.

In the ultrastrong coupling regime the above questions, e.g., how to set up tight-binding models, have attracted a lot of research interest lately \cite{Zueco2021b,rokaj_free_2021,li_electromagnetic_2020,dmytruk_gauge_2021,Ashida2021}. In this limit, solving the full problem of a light-matter coupled particle in a periodic potential proves more difficult. In addition to the Peierls substitution in Eq.~(2), there are dipolar and higher order matrix elements in the quantum light-matter Hamiltonian in a tight-binding basis \cite{Zueco2021b,li_electromagnetic_2020,dmytruk_gauge_2021}, and the validity of few-band approximations becomes increasingly questionable at ultrastrong coupling \cite{Ashida2021}. One strategy in the limit where light-matter interaction becomes the dominant energy scale was put forward in Ref.~\onlinecite{Ashida2021} and relies on a unitary transformation that achieves asymptotic decoupling of light and matter degrees of freedom.

\subsection{Cavity control of collective phenomena in quantum materials}
\label{subsec:collective}

\paragraph{Cavity superconductivity}

Some of the holiest grails in the field of quantum materials are the quests for higher-temperature superconductivity and for novel unconventional order parameters, in particular topological superconductivity. The possibility to reach higher critical temperatures by laser driving was suggested by a series of experiments indicating light-induced superconducting-like states with lifetimes of a few picoseconds \cite{cavalleri_photo-induced_2018}, recently extended to nanoseconds \cite{Budden_metastable2021}. There has been a plethora of possible theoretical explanations for these observations, and it is a fair statement that no final conclusion has been reached yet, neither on the nature of the light-induced states, nor on the microscopic mechanisms behind them. Nevertheless, theorists have been inspired by those developments to conceive ideas that might enable cavity superconductivity.   

In general, there are two main strategies to achieve this goal. The first one is to employ the direct coupling of electrons and photons and to use the cavity photons of mediators of an attractive force that gives rise to Cooper pairing and ultimately superconductivity. The second strategy is to couple the photons to other degrees of freedom (e.g., phonons, magnons, plasmons, excitons, or even a combination thereof), that in turn couple to the relevant electronic quasiparticles to provide a pairing mechanism. 

(i) Direct photon-mediated pairing. The most straightforward fashion in which coupling to cavity modes can impact superconducting pairing is by providing a pairing glue through virtual-photon exchange between electrons. This is in direct analogy to the conventional phonon-mediated pairing in BCS superconductors. The relevant coupling term for a dark cavity is the paramagnetic $\bm{A}\cdot\bm{p}$ term here. As noted in Ref.~\onlinecite{schlawin_cavity-mediated_2019}, the effective electron-electron interaction mediated by cavity photons is typically long-ranged in real space and therefore limited in momentum space to regions very close to the Fermi surface. This is indeed akin to forward scattering channels in superconductors \cite{rademaker_enhanced_2016}. 
%Therefore, Cooper pairs with constituent electrons at very different momenta across the Fermi surface are not directly coupled via such a short-momentum-range interaction. The upshot from this line of thought is that a pairing contribution from forward scattering is largely agnostic with regard to the overall pairing symmetry, which depends on momenta of electrons in Cooper pairs across the entire Fermi surface. 
In stark contrast to phonon-mediated pairing, however, interactions mediated by the paramagnetic coupling are of current-current type. Thus, according to Amp\`ere's law it is attractive for electrons residing on the same side of the Fermi surface, giving rise to a pair-density-wave superconducting instability.
Therefore, while the bare critical temperature that can be reached if only the photon-mediated contribution is taken into account in a dark cavity might be limited to the sub-Kelvin regime for realistic settings, cavity-mediated pairing can boost superconductivity stemming from other -- in particular unconventional -- pairing mechanisms.

\begin{figure}[t]
\centering
\includegraphics[width=\columnwidth]{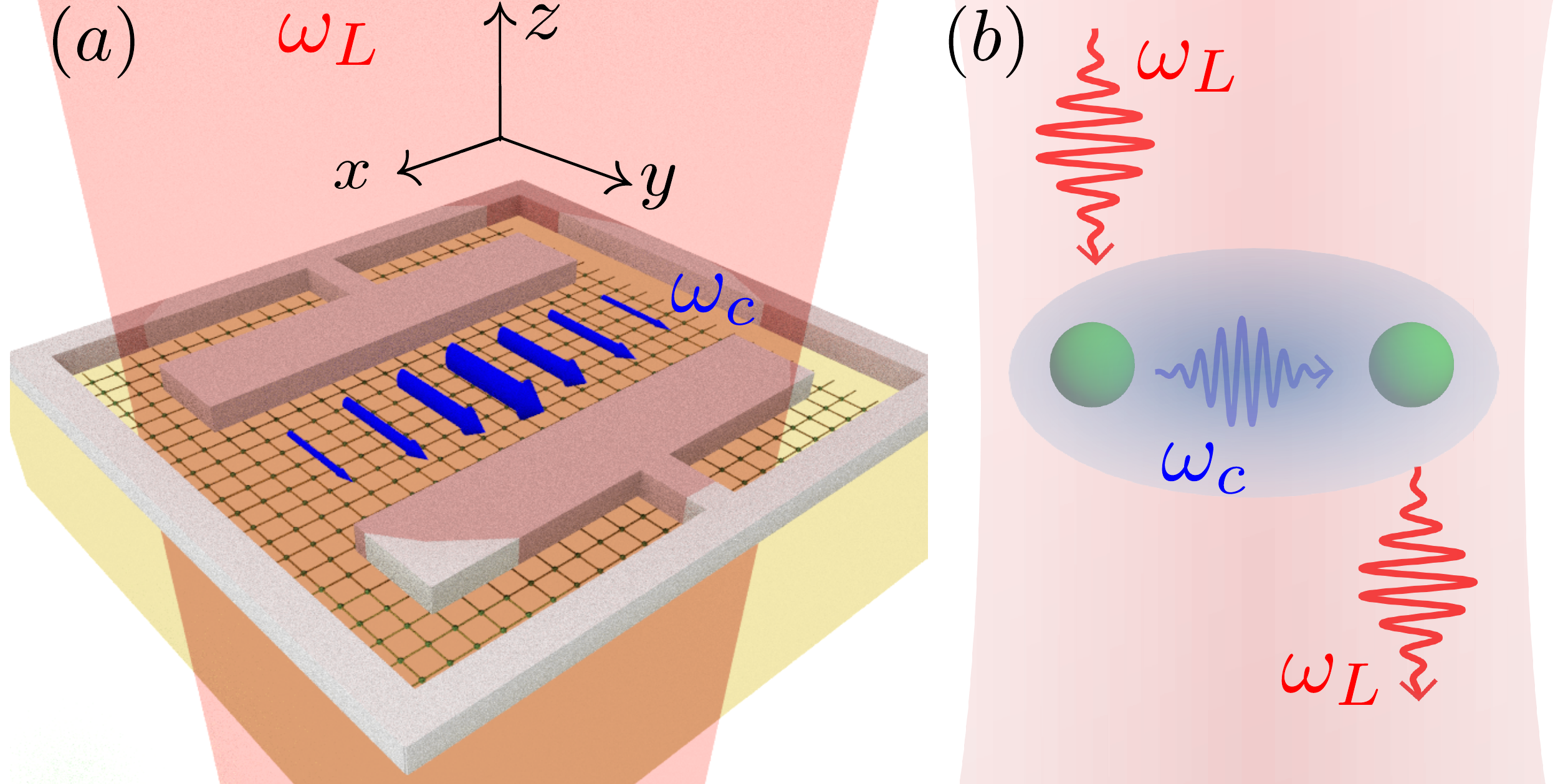}
\caption{
\textbf{One possible mechanism for photon-mediated electronic pairing in a driven cavity.} (a) Setup with a two-dimensional material (square lattice) coupled to a split-ring resonator with electric field indicated by blue arrows and resonance frequency $\omega_c$. The cavity-material system is driven by a laser field (red shading) at frequency $\omega_L$ detuned from $\omega_c$. (b) Two electrons (green spheres) with cavity-mediated interaction driven by diamagnetic electron-photon coupling with one laser photon (red wiggly arrow) and one cavity photon (blue wiggly arrow).
Figure reproduced with permission from Physical Review Letters 125, 053602 (2020), Ref.~\onlinecite{gao_photoinduced_2020}. Copyright 2020 American Physical Society.
}
\label{fig:cavity_sc}
\end{figure}

Another theoretical idea is to focus on the diamagnetic $A^2$ light-matter coupling term \cite{gao_photoinduced_2020,chakraborty_long-range_2021}. %Here, the basic principle is that one takes one photon in this two-photon term from an external driving laser and the second photon from the cavity.
Here, the basic idea is illustrated in Fig.~\ref{fig:cavity_sc}: Upon driving the cavity (here illustrated by a split-ring cavity) with an external laser (red), laser photons can scatter into and out of the cavity due to the diamagnetic interaction. 
In terms of the quantum photon degrees of freedom, this creates an effective one-photon coupling between the electrons and the cavity mode, with a prefactor that is determined by the laser amplitude. 
Laser photons can then virtually scatter into and out of the cavity via the two-electron process depicted in Fig.~\ref{fig:cavity_sc}(b), giving rise to an effective electron interaction. For this process to be effective, laser and cavity frequencies need to be sufficiently strongly detuned. Closer to resonance, the depicted interaction is strongly screened by plasmons and the induced interaction suppressed.
This type of pairing, which is closely related to well-established schemes for inducing long-range interactions in ultracold atoms~\cite{Ritsch2013, Mivehvar2021}, has two main advantages: (i) The strength of interaction can be tuned by an external laser drive; (ii) the sign of the interaction can be controlled by the laser frequency detuning with respect to the cavity resonance frequency. Another difference with respect to the coupling via the paramagnetic light-matter coupling lies in the fact that the diamagnetic coupling involves the electronic density (the kinetic energy density in tight-binding models) rather than the current density in the case of the paramagnetic term. Consequently, this scheme could enhance existing conventional pairing instabilities in quantum materials. Notably, the cavity-mediated coupling can be enhanced by an additional photon fluctuation contribution \cite{chakraborty_long-range_2021}, which is suppressed in conventional phonon-mediated pairing. We finally note that similar effects in excitonic systems were recently demonstrated in experiments \cite{Cortese2019, Cortese2021}. Here, the strong coupling to an ionising transition creates excitons bound by photon exchange.

\begin{figure}[t]
\centering
\includegraphics[width=\columnwidth]{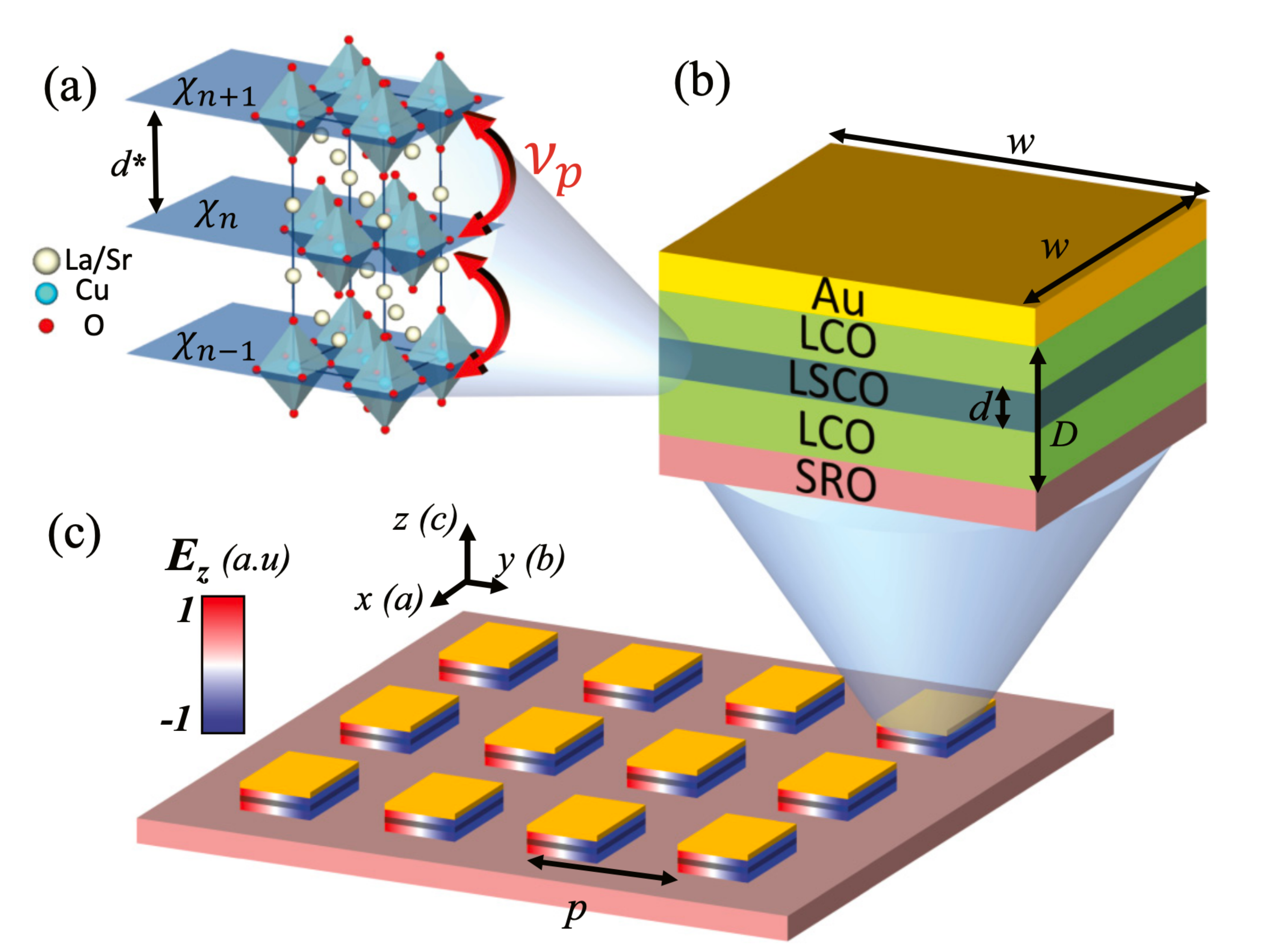}
\caption{
\textbf{Josephson plasmon polaritons.} (a) Sketch of the layered superconductor LSCO, which forms the active material in the proposal. 
(b) Integrated microcavity design composed of gold (Au) and SrRu$_3$ (SRO) mirrors, as well as La$_2$CuO$_4$ dielectric spacer material surrounding the La$_{2-x}$Sr$_x$Cu)$_4$ (LSCO) sample.
(c) The proposed heterostructure of several microstructured cavities. 
Figure reproduced with permission from Physical Review B 93, 075152 (2016), Ref.~\onlinecite{Laplace2016}. Copyright 2016 American Physical Society.
}
\label{fig:Josephson-plasmons}
\end{figure}

The long-range nature of this cavity-mediated interaction has important consequences for collective modes in the resulting superconducting state~\cite{Gao_Higgs2021}. In particular, the Higgs amplitude mode is pushed in-gap and Bardasis-Schrieffer exciton modes can be stabilized. In other works, superconductors in a cavity have also been predicted to host new or cavity-modified collective modes, for instance polaritons made out of Bardasis-Schrieffer excitons and photons \cite{allocca_cavity_2019}, or hybrid modes of the Higgs amplitude mode and cavity photons, coined cavity Higgs polaritons \cite{raines_cavity_2020}. In layered superconductors such as the cuprates, cavities may also be made to hybridize with Josephson plasmon excitations \cite{Laplace2016}. The latter proposal is shown in Fig.~\ref{fig:Josephson-plasmons}, where the superconducting material is integrated into a microcavity design. The lengths $w$ are chosen to harbour a $\lambda/2$ standing wave matched to the Josephson plasmon resonance. This hybridization was recently observed in the coupling of the plasmons with a split ring cavity in Ref.~\onlinecite{schalch_strong_2019}.
The strong coupling to a graphene plasmon cavity is predicted to reveal hyperbolic Cooper pair polaritons in cuprate superconductors, which could provide insights into unconventional superconductivity \cite{Berkowitz2021}.

The cavity coupling is also predicted to give rise to a quantum analogue of the Eliashberg effect \cite{curtis_cavity_2019}. In this quantum Eliashberg effect, the fluctuations of a strongly coupled electromagnetic field can redistribute electronic quasiparticles and thereby enhance the critical temperature.

\begin{figure*}[t]
\centering
\includegraphics[width=\textwidth]{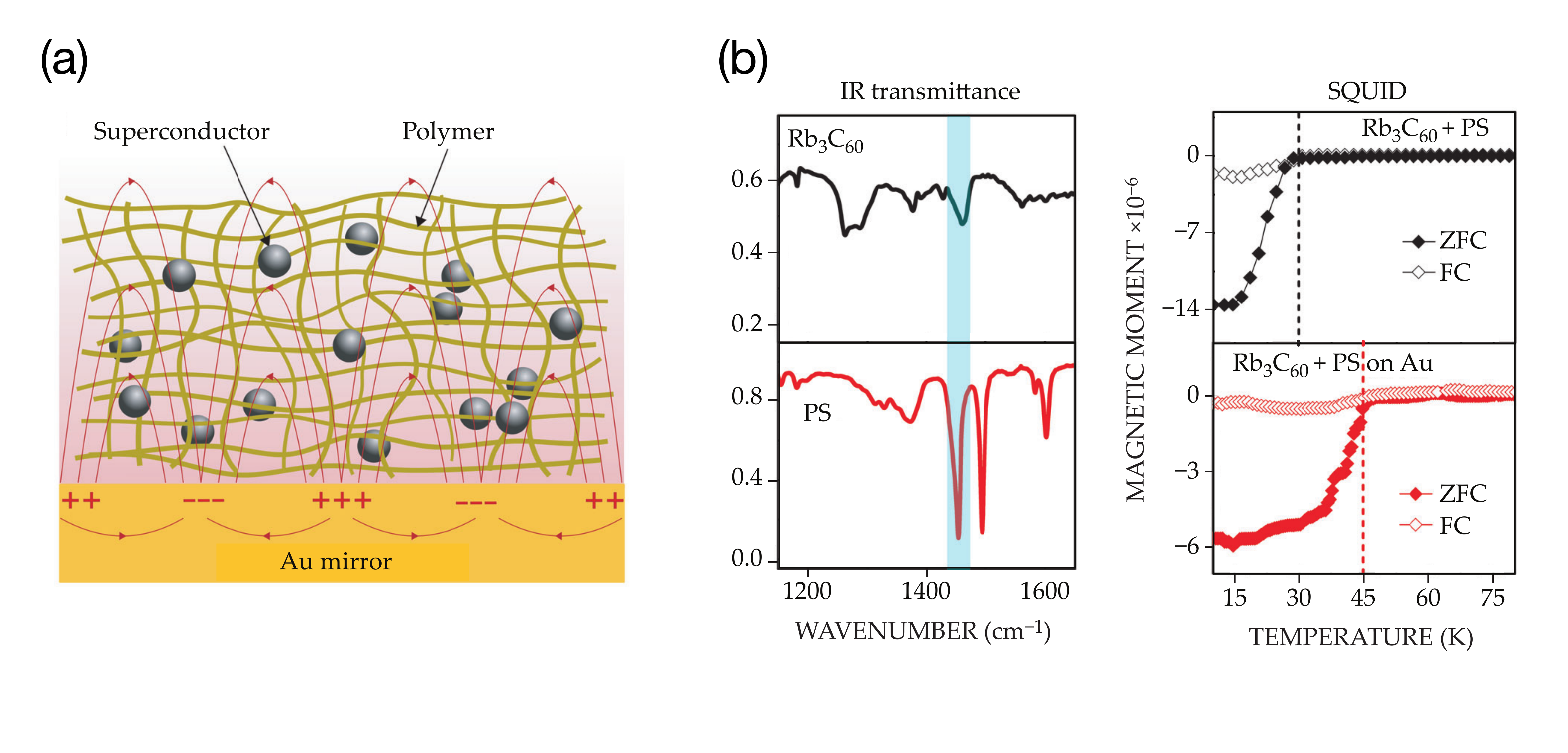}
\vspace{-1.7cm}
\caption{
\textbf{Light-enhanced superconductivity through a plasmonic surface.} (a) Superconductor (Rb$_3$C$_{60}$) inside a polymer matrix placed on top of a gold surface. The polymer's purpose is to mediate a coupling between surface plasmon polaritons of gold and vibrational modes in the superconductor through its infrared-active phonons. (b) The infrared (IR) transmittance of Rb$_3$C$_{60}$ is only weakly suppressed between 1400 and 1500 $cm^{-1}$, pointing to a weak IR absorption in this wavenumber range (upper left panel). By contrast, polystyrene (PS) has a strong IR absorption in this regime through its vibrational modes (lower left panel). Right panels: SQUID measurements of the superconducting-polymer system's magnetic moment under zero-field-cooled (ZFC) and field-cooled (FC) conditions suggest an increase in the superconducting critical temperature from 30 K to 45 K.
Figure reproduced from Physics Today 74, 42-48 (2021), Ref.~\onlinecite{genet_inducing_2021}, with the permission of AIP Publishing; adapted from Ref.~\onlinecite{thomas_exploring_2019}.
}
\label{fig:ebbesen}
\end{figure*}

(ii) Indirect impact on pairing mechanism via other collective modes. As discussed above, the direct coupling of photons to electrons via minimal coupling (quasi-free electrons) or the Peierls substitution (tight-binding electrons) is often relatively weak. However, photons can couple to other degrees of freedom that in turn affect the pairing in a superconductor. 

Most notably, phonons are known to be key for phonon-mediated superconductivity but also play a role in high-temperature superconductors. The basic idea to employ phonon polariton formation in order to modify the effective electron-phonon coupling in a superconductor was suggested in Ref.~\onlinecite{sentef_cavity_2019}. In this work, coupling of a branch of cavity modes to the out-of-plane optical polar phonon mode at the interface between monolayer FeSe and a SrTiO$_3$ substrate was considered together with the forward-scattering electron-phonon interaction to the quasi-two-dimensional conduction electrons in FeSe. This results in an indirect electron-photon coupling via phonon polaritons, and direct electron-photon coupling was neglected since the polarization of the photon mode is orthogonal to the electronic current in that geometry, such that the paramagnetic coupling vanishes, and only a weaker diamagnetic electron-photon coupling is present. A diagrammatic Migdal-Eliashberg treatment of the resulting electron-phonon-photon system then revealed that the softening of the lower polariton branch compared to the bare phonon mode then indeed leads to an enhanced effective electron-phonon coupling strength (parametrized by the electronic self-energy and the related electronic mass enhancement). However, due to the forward-scattering, small-momentum transfer nature of phonon-mediated superconductivity in that particular model, the resulting superconducting critical temperature was predicted to be decreased by the presence of the cavity. 

A follow-up theoretical work then employed a similar formalism involving surface plasmon polaritons and phonons to predict enhanced superconducting critical temperatures in more conventional BCS superconductors including larger momentum transfers between electrons and phonons \cite{hagenmuller_enhancement_2019}. A possible experimental confirmation of the plasmon- and phonon-mediated mechanism of superconductivity enhancement was then reported in Ref.~\onlinecite{thomas_exploring_2019} (also see discussion in Ref.~\onlinecite{genet_inducing_2021}).
The setup employed in this work is sketched in Fig.~\ref{fig:ebbesen}(a). 
Grains of superconducting material (Rb$_3$C$_{60}$) are embedded in a polymer (polystyrene) matrix and deposited on a gold surface. The key feature of the polymer is that it strongly absorbs in the infrared and can thus, presumably, act as a mediator of coupling between the superconductor's optical phonons and the surface-plasmon polaritons on the gold surface (Fig.~\ref{fig:ebbesen}(b), left panel).
Remarkably, magnetometry measurements with a superconducting quantum interference device (SQUID) showed a drop in the magnetic moment at an enhanced critical temperature of 45 K compared to 30 K without the gold surface (Fig.~\ref{fig:ebbesen}(b), right panel)
, together with a characteristic deviation between field-cooled and zero-field-cooled curves, indicative of a Meissner effect. The mediator mechanism of the polymer was tested experimentally by replacing polystyrene with other polymers with different infrared absorption spectrum, in which case no effect of the surrounding on the superconductivity was observed. Taken together, this evidence suggests that superconductivity can indeed be surprisingly strongly affected by cavity-like surroundings in conjunction with phonon polariton formation. Clearly, further experimental work exploring this new field of cavity superconductivity and clarifying the microscopic mechanism is desirable.

In related works, exciton polaritons \cite{laussy_exciton-polariton_2010,cotlet_superconductivity_2016,Kavokin2016} or hyperbolic plasmons \cite{grankin_interplay_2022} were considered as potential mediators of cavity superconductivity. Moreover the squeezing of phonons involved in superconducting pairing was suggested as another route to enhancing superconductivity \cite{grankin_enhancement_2021}, in close similarity to earlier proposals of electronic squeezing of classically driven phonons \cite{Kennes2017}. Finally, motivated by the observation that classical laser driving of specific phonon modes that modulate electron-electron interactions leads to transient superconducting-like states in organic molecular crystals \cite{buzzi_photomolecular_2020}, the influence of phonon polariton formation involving such modes in cavities on the effective electron-electron interactions was recently studied \cite{le_de_cavity_2022}. We refer to an overview of the zoo of polaritonic modes that are known, which could be considered as candidates to bring forward similar ideas, for further reading \cite{basov_polariton_2021}. 

\paragraph{Cavity phononics and ferroelectricity}

The formation of cavity polaritons is one obvious consequence of the coupling of photonic degrees of freedom to collective modes in a material, e.g., to phonons. Therefore, it is a natural question to ask whether the presence of a cavity can also lead to softening and ultimately instability of the crystal lattice towards ferroelectric ordering. This has been explored theoretically both in general model settings \cite{ashida_quantum_2020} and with ab initio inputs with an eye towards SrTiO$_3$ \cite{latini_ferroelectric_2021}. The latter material is of particular interest since its metastable ferroelectricity via optical straining with terahertz fields has been demonstrated experimentally \cite{li_terahertz_2019,nova_metastable_2019} and investigated theoretically \cite{shin_simulating_2021}. 

\begin{figure}[t]
\centering
\includegraphics[width=\columnwidth]{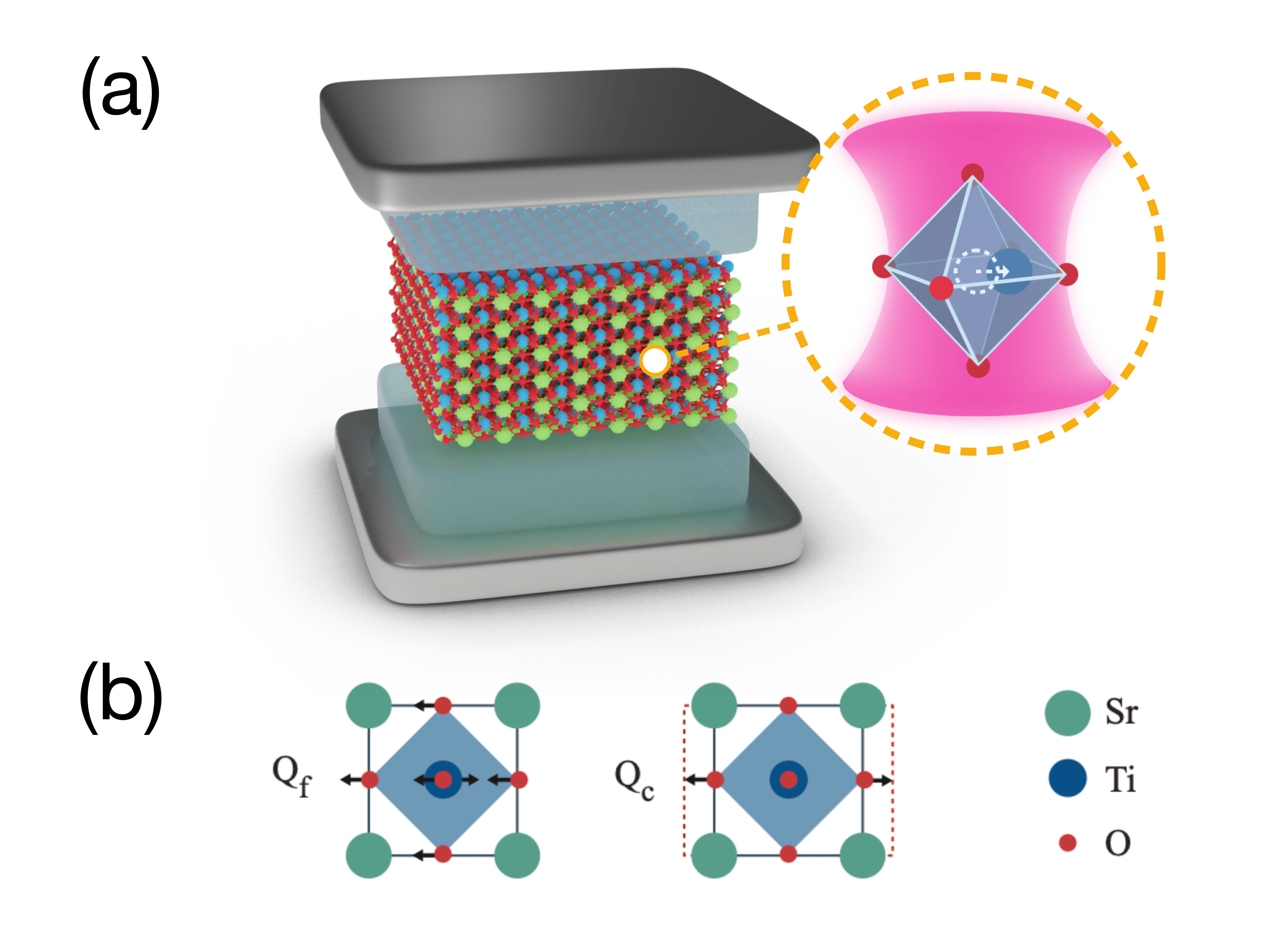}
\caption{
\textbf{Ferroelectricity mediated by a dark optical cavity.} (a) Cartoon of SrTiO$_3$ inside a dielectric medium in a Fabry-P\'erot cavity. The zoom bubble highlights a Ti-O displacement induced by the photons in the cavity. (b) Relevant lattice displacements involved in ferroelectric symmetry breaking: the ferroelectric soft mode with coordinate $Q_f$ and a second lattice vibration with coordinate $Q_c$, respectively. 
Figure taken from S. Latini, D. Shin, S. A. Sato, C. Schäfer, U. D. Giovannini, H. Hübener,
and A. Rubio, Proceedings of the National Academy of Sciences 118, e2105618118 (2021), Ref.~\onlinecite{latini_ferroelectric_2021}; licensed under a Creative Commons Attribution (CC BY) license.
}
\label{fig:cavity_ferro}
\end{figure}

Consider a dark optical cavity with a sample of SrTiO$_3$ embedded in a dark optical cavity with a dielectric medium (Fig.~\ref{fig:cavity_ferro}(a)). A ferroelectric state is formed by a relative displacement between Ti and O atoms. The relevant ferroelectric phonon mode with coordinate $Q_f$  is coupled to both the photon field $\hat{A} \propto (a^\dagger + a)$ as well a to a second phonon mode $Q_f$ (Fig.~\ref{fig:cavity_ferro}(b)). The theoretical prediction by Latini et al., Ref.~\onlinecite{latini_ferroelectric_2021} is that the combined hybridization between ferroelectric mode and the photon and second phonon modes leads, above a critical light-matter coupling strength, to a softening of the ferroelectric mode indicating the instability of the system to ferroelectricity. This was coined ``ferroelectric photo ground state''. In the closely related work of Ashida et al., Ref.~\onlinecite{ashida_quantum_2020} different couplings involving photons, phonons and plasmons in paraelectric systems between cavity mirrors were analyzed, and analogies between the ferroelectric phase and a superradiant phase transition (see below) were pointed out. Also a cavity extension of nonlinear phononics \cite{disa_engineering_2021} was suggested \cite{juraschek_cavity_2021}.

\begin{figure}[t]
\centering
\includegraphics[width=\columnwidth]{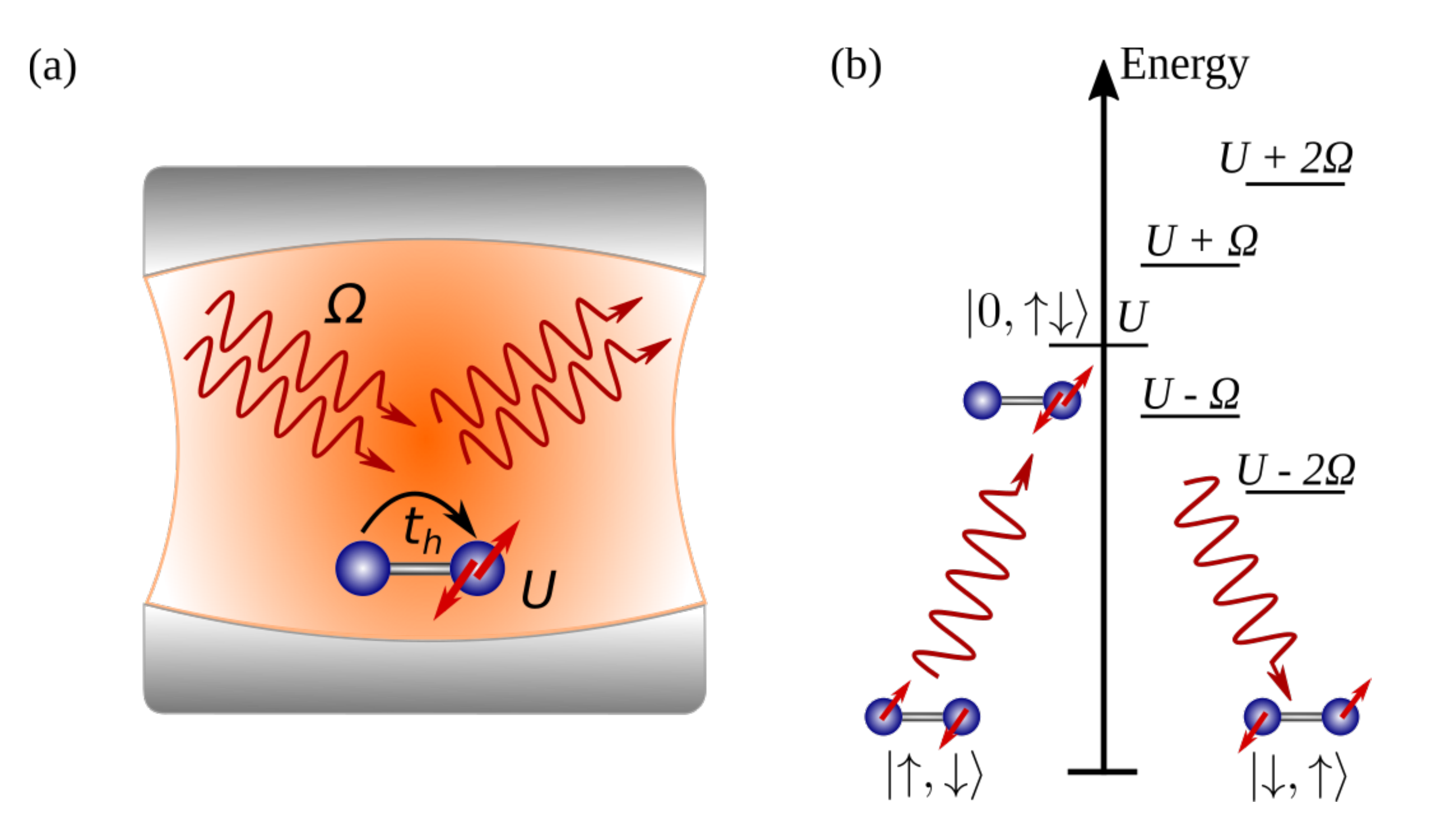}
\caption{
\textbf{Hubbard model coupled to a photon mode.} (a) Sketch of Hubbard dimer with hopping $t_h$ and on-site interaction $U$ coupled to a cavity photon mode with frequency $\Omega$. (b) Energetics of relevant virtual processes leading to cavity-modified spin exchange interaction in the $U\gg t_h$ limit. Spins on neighboring lattice sites are exchanged through virtual processes involving doubly occupied sites and different numbers of photon-absorbed and photon-emitted states with energies $U + \ell \Omega$ as indicated.  
Figure taken from M. A. Sentef, J. Li, F. Künzel, and M. Eckstein, Physical Review Research 2, 033033 (2020); Ref.~\onlinecite{sentef_quantum_2020}; licensed under a Creative Commons Attribution (CC BY) license.
}
\label{fig:cavity_hubbard}
\end{figure}

\paragraph{Correlated systems in a cavity}
An entirely new research field is opened when one bridges strongly correlated electrons and their coupling to a cavity. Starting from models of itinerant electrons with strong correlations, general considerations have shown how electron-photon coupling can lead to modified effective spin-exchange interactions 
\cite{kiffner_manipulating_2019,kiffner_mott_2019}. Extending these ideas to driven cavities, the quantum-to-classical crossover of Floquet engineering of magnetic kinetic exchange interactions in Peierls-coupled Hubbard models downfolded to cavity spin models was discussed \cite{sentef_quantum_2020}. To understand the basic physics, we briefly describe the cavity Hubbard model (Fig.~\ref{fig:cavity_hubbard}(a))
\begin{align}
    \hat{H} =& t_{h} \sum_{j\sigma} \left( \hat{c}_{j,\sigma}^\dagger \hat{c}_{j+1,\sigma}^{} \; e^{i\hat{A}} + \textrm{H.c.} \right) + 
    \nonumber \\ &+
     U \sum_j \hat{n}_{j,\uparrow} \hat{n}_{j,\downarrow} + \Omega \hat{a}^{\dagger} \hat{a}^{}
    \label{eq:cavity_hubbard}
\end{align}
with local Hubbard interaction $U$, $t_h$ is the electronic hopping matrix element between neighboring atoms, and $\Omega$ denotes the bare cavity photon frequency. Here the cavity photon vector potential is $\hat{A} = g \left( \hat{a}^{} + \hat{a}^{\dagger} \right)$, with $g$ a dimensionless light-matter coupling strength determined by the cavity setup, and the operators $a,a^\dag$ annihilate/create photons. The electronic annihilation (creation) operators $c_{j,\sigma}^{(\dagger)}$ acting on site $j$ and spin $\sigma=\uparrow,\downarrow$ have number operators $\hat{n}_{j,\sigma} = \hat{c}_{j,\sigma}^\dagger \hat{c}_{j,\sigma}^{}$. In Ref.~\onlinecite{sentef_quantum_2020} it was shown that a suitable cavity Schrieffer-Wolff transformation leads to a cavity-renormalized effective Heisenberg model in the strong-coupling $U \gg t_h$ limit, which for neighboring sites $1$ and $2$ reads
\begin{align}
\label{eq:cavity_heisenberg}
H = (\bm S_1 \bm S_2 -\tfrac12)\mathcal{J}[a^\dagger,a] + \Omega \hat{a}^\dagger \hat{a}.
\end{align}
Here $\bm S_i$ is the usual spin-$\frac12$ operator on site $i$, and $\mathcal{J}[\hat{a}^\dagger,\hat{a}]$ an operator acting on the photon states, which reduces to the bare spin-exchange interaction $J_{\rm ex}$ in the absence of the cavity. This can be contrasted with the classical Floquet result of a Hubbard system driven with a coherent field amplitude $E = A / \Omega$ \cite{mentink_ultrafast_2015}
\begin{align}
J^{F}_{\rm ex}=J_{\rm ex}\sum_{\ell}\frac{J_{|\ell|}(A)^2}{1+\ell \Omega/U},
\label{eq:classical_floquet_exchange}
\end{align}
with $J_{\rm ex}=4t_h^2/U$ the exchange interaction of the undriven system and $J_{\ell}(A)$ the $\ell$th Bessel function. Both in the classical and quantum limits the photon-dressed spin exchange features contributions from virtual processes involving doubly occupied intermediate states with $\ell$ photons leading to energy denominators $U + \ell \Omega$ (Fig.~\ref{fig:cavity_hubbard}(b)). The key difference between quantum and classical cases is that the probabilities of photon emission and absorption become unequal in the quantum limit and depend on the actual photonic wavefunction in the coupled cavity-matter system. In Ref.~\onlinecite{sentef_quantum_2020} it was further shown that already few-photon states in a cavity lead to renormalization of the exchange coupling in close similarity to the classical Floquet case provided that the light-matter coupling is strong enough. This result implies that (i) coherence of photons is not a requirement for Floquet effects in the strong-coupling regime, and (ii) that strong laser driving might not be necessary. 
In a ground state cavity, these renormalization effects reduce magnetic interacions and compete with long-range interactions mediated by the cavity, which are analogous to those discussed above in superconductors \cite{kiffner_manipulating_2019}.
Optically bright excitations in the Hubbard model can further hybridize to create Mott polaritons, thus strongly affecting the optical conductivity of the coupled cavity-electron system \cite{kiffner_mott_2019}.

An investigation of a system with intertwined charge-density wave and superconducting orders (attractive $U$ Hubbard chain) showed how coupling to a cavity can selectively suppress or enhance the charge and pair correlations depending on the detuning between the cavity frequency and the Hubbard $U$ parameter \cite{li_manipulating_2020}. Starting from a two-dimensional Heisenberg-type spin model with short-range spin exchange and long-range dipolar interactions, it was shown how coupling of the electrons to cavity photons can lead to cavity-induced quantum spin liquid phases even in originally unfrustrated spin systems
\cite{chiocchetta_cavity-induced_2021}. A recent study investigated opportunities for coupling spin excitations (magnons) relevant for parent compounds of high-temperature cuprate superconductors to cavity modes through spin-orbit coupling and the crystal lattice, paving the way for ideas to employ the hybridization between spin fluctuations in doped materials and cavity photons to potentially affect high-temperature superconductivity
\cite{curtis_cavity_2021}.
Finally, it was shown that the coupling of a cavity to interband transitions could enhance an instability towards the formation of an excitonic insulator \cite{mazza_superradiant_2019}. This excitonic condensation takes place concomitantly with a superradiant phase transition in the cavity which we will discuss below. A strong enhancement of ferromagnetism in YBCO nanoparticles was reported in recent experiments \cite{thomas_large_2021}, employing light-matter collective strong coupling involving gold mirrors and polystyrene embeddings, similar as in Ref.~\onlinecite{thomas_exploring_2019}.

\paragraph{Light-magnon coupling} Strong, single-particle (photon/magnon) coherent coupling between the magnetic long-wavelength modes of a magnetically ordered
magnetic material and the electromagnetic modes of a microwave cavity was theoretically predicted in 2010 by Soykal and Flatt\'e \cite{soykalStrongFieldInteractions2010b}.
Seminal experimental papers demonstrated this coupling regime \cite{hueblHighCooperativityCoupled2013a,tabuchiHybridizingFerromagneticMagnons2014b,zhangStronglyCoupledMagnons2014b,goryachevHighCooperativityCavityQED2014}. The employed material is usually yttrium iron garnet (Y$_{3}$Fe$_{5}$O$_{12}$, YIG), an insulator known for its low Gilbert damping, which governs the dissipation of the magnon modes. In this regime of GHz frequencies the magnetization of the sample couples to the magnetic field of the microwave (MW) mode via the usual dipolar Zeeman coupling, resulting in a coupling
of the type $g_{\text{MW}}(m^{\dagger}+m)(a^{\dagger}+a)$ between magnons $m^{(\dagger)}$ and photons $a^{(\dagger)}$. The strong coupling is reflected in the hybridization of the microwave and magnon modes at resonance, with very high cooperativities (e.g., $10^{7}$, see ~Ref.~\onlinecite{goryachevHighCooperativityCavityQED2014}). The resonance, denominated a cavity magnon polariton, can be tuned by an external magnetic field which controls the frequency of the magnon modes. MW cavities have been used to mediate strong coherent coupling between magnons and
a superconducting qubit \cite{tabuchiCoherentCouplingFerromagnetic2015b}.
%\textcolor{red}{{[}OPTIONAL REF Single shot single-magnon detection
%has been demonstrated using this setup and spectroscopic measurements
%of the qubit \cite{lachance-quirionEntanglementbasedSingleshotDetection2020}{]}.}
%\textcolor{red}{{[}OPTIONAL, BUT THESE ARE CURRENTLY IMPORTANT DIRECTIONS
%IN THE FIELD
The tunability of these systems makes them amenable to
explore non-hermitian physics by designing the coupling to be in the dissipative regime \cite{wangDissipativeCouplingsCavity2020}. Moreover,
coherent coupling of cavity magnon polaritons to phonons can be probed
and harnessed \cite{zhangCavityMagnomechanics2016,pottsDynamicalBackactionMagnomechanics2021a}.
We note that photon condensation and cavity-enhanced magnetism was also suggested theoretically recently \cite{Zueco2021}.

Whereas strong resonant coupling of magnons to photons in the
MW regime is by now routinely realized, achieving strong coupling to optical photons (usually infrared \textasciitilde{} 200 THz, where YIG is transparent), which are highly detuned, is much more challenging.
At these frequencies, the permeability is that of the vacuum and the magnetization couples perturbatively to the electric field. The resulting coupling is proportional to the optical spin density of the EM field and therefore quadratic in the electric field, $g_{\text{opt}}m^{\dagger}(a_{1}^{\dagger}a_{2})+h.c.$. The same coupling mechanism is responsible for magnetic Brillouin scattering and the Faraday effect, and its strength is proportional to the Verdet constant of the material. YIG has a reasonably large Verdet constant, which is why it is used for Faraday rotators and isolators. Seminal experiments
demonstrated coherent coupling in YIG by using an optical cavity to enhance the coupling \cite{haigh_triple-resonant_2016,zhang_optomagnonic_2016,osadaCavityOptomagnonicsSpinOrbit2016}.
The cavity is in this case realized by the material itself when properly patterned -- the light is trapped by total internal reflection. The typical geometry is a sphere of YIG hosting optical whispering gallery modes. Although the theoretically predicted optimal coupling is large
\cite{violakusminskiyCoupledSpinlightDynamics2016,liuOptomagnonicsMagneticSolids2016a}
(\textasciitilde MHz, comparable to the coupling in optomechanics
where optical photons couple to phonons), in practice the coupling remains weak due to poor mode overlap and large mode volumes. 
%\textcolor{red}{{[}OPTIONAL
%REFs Nevertheless progress is being made in terms of the size of the
%cavity\cite{haighSubpicoliterMagnetopticalCavities2020}. A promising
%route for achieveing strong coupling between magnons and photons in
%the optical regime could be realized by using optomagnonic crystals:
%a periodically patterned magnetic dielectric with a defect which co-localizes
%magnon and photon modes in a small, micron sized volume \cite{grafDesignOptomagnonicCrystal2021}.
%Realizing strong coupling also in the optical regime could open the
%door for quantum transduction using magnons. {]}}

\paragraph{Topology and quantum Hall effect}
%Faist experiment
%MICHAEL: Faist figure cannot be reproduced

Cavity quantum Hall systems form a very active field of research since the first demonstration of many-body quantum optics with Landau polaritons \cite{scalari_ultrastrong_2012} - hybridisations of a cavity mode with the cyclotron transition of a two-dimensional electron gas. We have mentioned several seminal experiments from this field already in this review, including the observation of high cooperativities in solid state cavity QED \cite{Kono2016} or of the Bloch-Siegert shift \cite{Kono2018}. Other notable examples include the change of Shubnikov-de-Haas oscillations in transport measurements \cite{paravicini-bagliani_magneto-transport_2019} and the impact of a nonparabolic bandstructure \cite{keller_landau_2020}.
%In related works the hybridization of cavity modes with Landau levels of two-dimensional electron gases in magnetic fields, so-called Landau polaritons, was demonstrated experimentally \cite{paravicini-bagliani_magneto-transport_2019,keller_landau_2020}.

Floquet engineering of topology was started by the insight that circularly polarized light breaks time-reversal symmetry and therefore opens a gap at a two-dimensional Dirac point \cite{oka_photovoltaic_2009,kitagawa_transport_2011}; this led to the research field of Floquet topological states of matter \cite{lindner_floquet_2011}. Since the key ingredient behind the gap opening is not the presence of a quasi-classical coherent state of many photons but rather the time-reversal symmetry breaking through circular light polarization itself, it is natural to consider quantum fluctuations of a circularly polarized mode in a chiral cavity as a different way of opening the gap and inducing nontrivial topology. The gap opening through vacuum dressing of Dirac fermions with circularly polarized light was first discussed in Ref.~\onlinecite{kibis_band_2011} and later extended to spectral functions and the quantized anomalous Hall response in a cavity Chern insulator
\cite{wang_cavity_2019}, i.e., an insulator with a band gap and a nonzero Chern number due to broken time-reversal symmetry. 

We briefly discuss the key ingredients that lead to the cavity Chern insulator. The bare electronic structure is a branch of Dirac fermions with momentum-dependent $2 \times 2$ Hamiltonian $h(\bm{k}) = \hbar v_F (k_x \sigma_x - k_y \sigma_y)$, with Fermi velocity $v_F$ and Pauli matrices $\sigma_i$. The Dirac fermion is coupled via $\hbar\bm{k} \rightarrow \hbar\bm{k}-e\bm{\hat{A}}$ to the photonic vector potential $\bm{\hat{A}}$. For right-handed circularly polarized light with polarization vector $\bm{e} = \frac{1}{\sqrt2} (1,i)$, this leads to the replacement $\hbar v_F (k_x + i k_y) - e A_0 \sqrt{2} a^\dagger$ in $h(\bm{k})$, where $a^\dagger$ creates a photon in the lowest cavity resonance mode with frequency $\omega$, and $A_0 = \sqrt{\hbar/(\epsilon \epsilon_0 V \omega)}$, with vacuum permittivity $\epsilon_0$, cavity volume $V$, and dielectric constant $\epsilon$ of the dielectric embedding of the two-dimensional material in the cavity. In Ref.~\onlinecite{wang_cavity_2019} a second-order many-body perturbation theory calculation of the cavity-induced electronic self-energy was performed, resulting in an intra-sublattice self-energy with the diagonal structure of the $\sigma_z$ Pauli matrix. At the Dirac point, a gap opening in the weak-coupling limit was predicted, with energy gap $\Delta = 2 \hbar^{-1}e^2 v_F^2 A_0^2/\omega$. Notably, this result is in exact analogy to the Floquet result in the high-frequency limit (or weak-driving) limit, with the only difference being that the amplitude of the classical circularly polarized laser vector potential $A_0$ (Floquet) is replaced here by the amplitude of the \emph{quantum fluctuations} of circularly polarized photon mode. We refer to  recent works on potential implementations and applications of chiral optical cavities \cite{hubener_engineering_2021} and chiral quantum optics \cite{petersen_chiral_2014,lodahl_chiral_2017,zhang_chiral_2019} for further reading.

While the proposed cavity-induced Chern insulator remains a challenge for future experiments, the opposite question has already been investigated, namely whether the presence of a cavity can affect and potentially destroy topological protection. To this end, a Hall bar with a two-dimensional (2D) electron gas was embedded in a terahertz split-ring resonator, and the corresponding longitudinal and transverse resistances measured as a function of an external magnetic field perpendicular to the sample plane \cite{appugliese_breakdown_2021}. It was shown that the integer quantum Hall plateaus were significantly affected by the presence of the cavity photon field, compared to the reference sample without resonator. These results were interpreted as a cavity-induced breakdown of topological protection due to cavity-mediated long-range hopping processes (also cf.~the related theory work Ref.~\onlinecite{ciuti_cavity-mediated_2021}). By contrast, the fractional quantum Hall plateaus were found to be largely unaffected by the cavity. The cavity modification of Hofstadter butterfly fractals in magnetotransport experiments was theoretically proposed in Ref.~\onlinecite{rokaj_polaritonic_2021}.

\begin{figure}[t]
\centering
\includegraphics[width=\columnwidth]{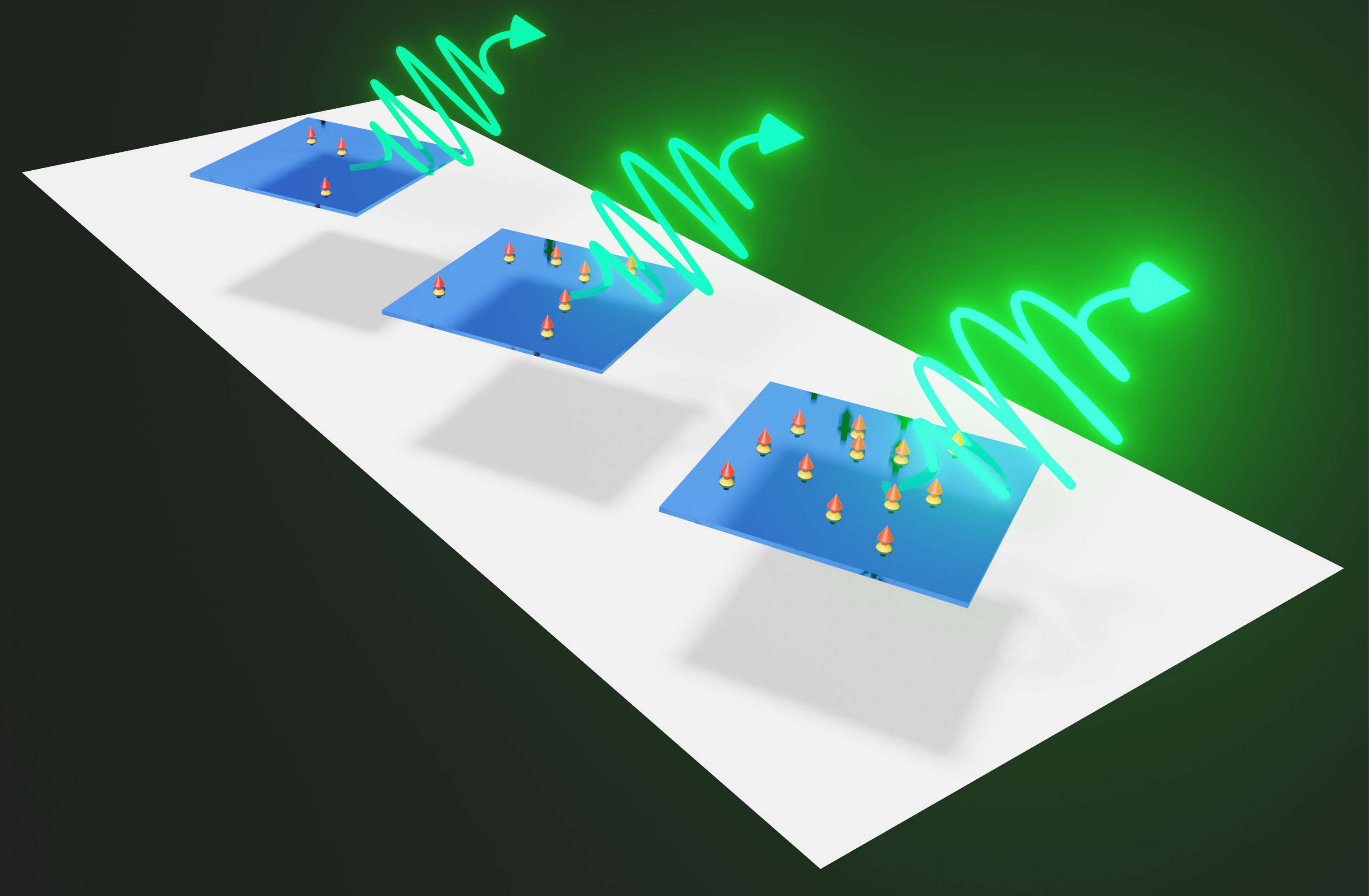}
\caption{
\textbf{Illustration of superradiance in the Dicke model.} 
The likeliness to emit a photon out of the excited states for an ensemble of identical two level systems increases with the number of its constituents. In the thermodynamic limit a spontaneous transition into a state with macroscopically occupied photon mode is found in the Dicke model. 
}
\label{fig.dicke}
\end{figure}

\paragraph{Superradiance}
In 1954 Dicke made the astonishing theoretical discovery that the probability of an excited atom (or two-level system) to emit a photon increases drastically when a second atom is placed nearby \cite{dicke_coherence_1954}. This puzzling effect even appears if the atoms are not directly coupled and even if the second atom is not in the excited but in its ground state. Dicke explained in his seminal work that this effect is driven by the mutual interaction of the atoms with a common light field. Under the assumption that the wavelength of the light is much larger than the separation of the two-level systems which are emitting the photons, these emitters can coherently interact with the light giving rise to a collective phenomenon called {\it superradiance} in the limit of many emitters. Superradiance leads to a collective emission of photons scaling as $N^2$ with the number $N$ of emitters. In a simplified Dicke model described by the Hamiltonian (neglecting terms $\mathcal{O}\!\left(A^2\right)$)
\begin{equation}
H=\hbar \omega_z \sum_{j=1}^N \sigma^z_j+\frac{2\lambda}{\sqrt{N}}(a+a^\dagger)\sum_{j=1}^N \sigma^x_j+\hbar\omega_c a^\dagger a
\end{equation}
superradiance manifests itself as a continuous transition at a critical coupling strength $\lambda_c$ at which the vector potential $A\sim a +a^\dagger$ spontaneously exhibits a macroscopic value $\left\langle a \right\rangle\sim \sqrt{N}$.  Experimental demonstrations \cite{skribanowitz_observation_1973,scheibner_superradiance_2007,timothy_noe_ii_giant_2012,baumann_dicke_2010} of a related effect in the nonequilibrium pumped and open Dicke model was associated with this equilibrium type of superradiance as well \cite{hepp_superradiant_1973}, but in its core is physically distinct from equilibrium superradiance.

%The discovery 
The prospect of superradiance has raised tremendous research attention. In the context of cavity materials \cite{schuler_vacua_2020} it has been conjectured that a similar effect might be found; after all a collection of two-level systems \cite{de_bernardis_cavity_2018} is not so different from spinful electrons moving in electronic bands. However, it was soon realized that keeping only the lowest order term linear in $A$ of a Peierls substitution \cite{eckhardt_quantum_2021} or only the linear order in $A$ for the electron gas \cite{rokaj_free_2021} breaks gauge invariance and can lead to a false superradiant phase transition, in close similarity to the Dicke model \cite{jaako_ultrastrong-coupling_2016,rzazewski_phase_1975,bialynicki-birula_no-go_1979,gawedzki_no-go_1981,gulacsi_floquet_2015}. False superradiant transitions can easily be found in such erroneous treatments and higher order terms restoring gauge invariance will provide no-go theorems \cite{nataf_no-go_2010,andolina_cavity_2019,stokes_uniqueness_2020} for the equilibrium superradiance transition in spatially homogeneous cavity fields. One way to circumvent these no-go theorems was suggested, namely taking into account spatially inhomogeneous modes \cite{nataf_rashba_2019,andolina_theory_2020,guerci_superradiant_2020} that transfer a finite momentum $\bm{q}$ to the electrons in an electron-photon scattering process. Moreover, whether the nonequilibrium version of superradiance can be found in quantum materials driven by laser pulses remains an intriguing question of current research (Fig.~\ref{fig.dicke}). 

\begin{figure*}[t]
\centering
\includegraphics[width=0.9\textwidth]{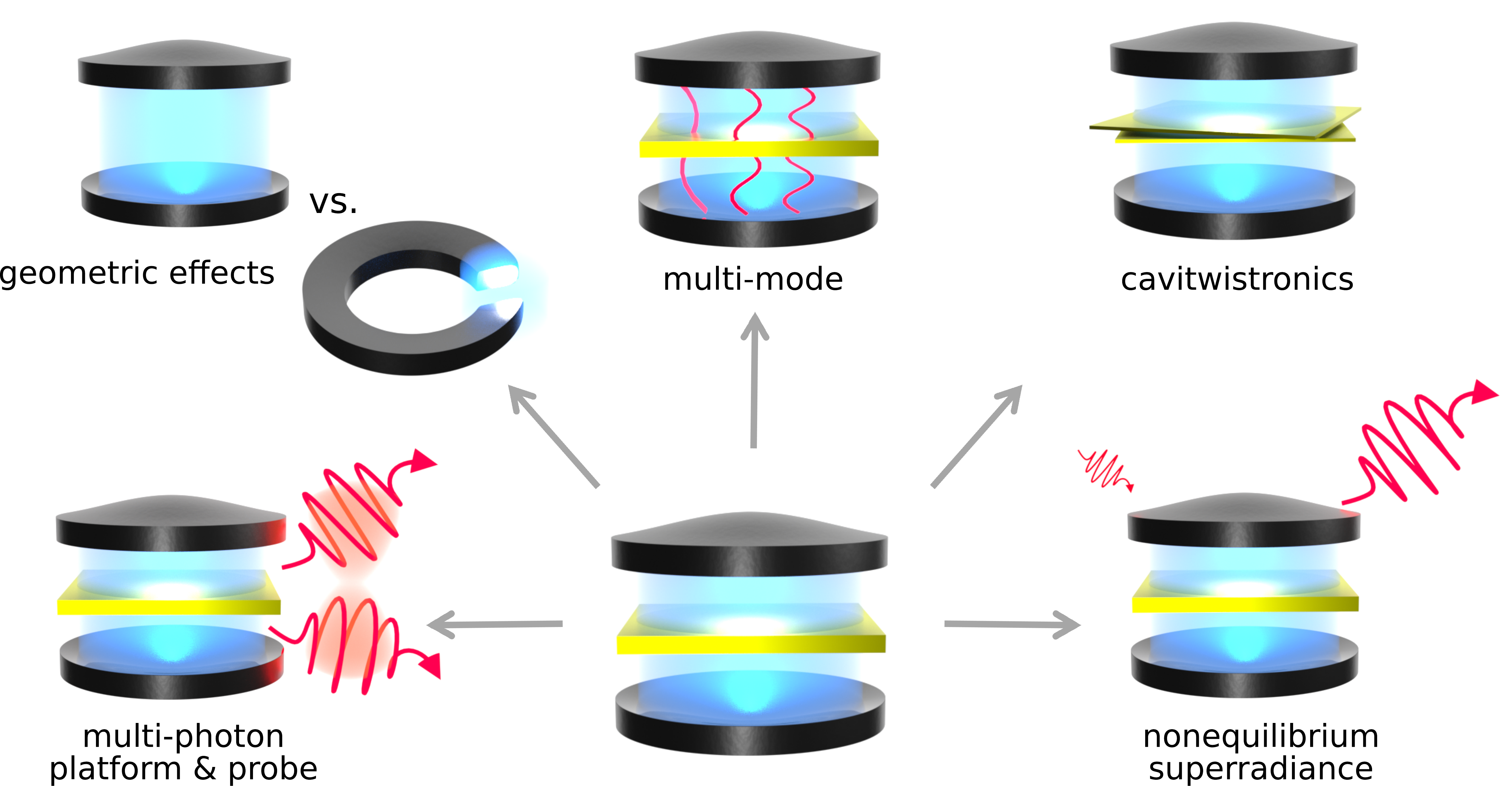}
\caption{
\textbf{Potential future directions in cavity quantum materials.} Here we summarize some future directions for the field of cavity quantum materials (center), expanding into five different avenues (arranged on the outer circle). On the level of modeling improvements in keeping multi-mode and geometric effects must be undertaken. Furthermore, in terms of quantum materials it would be intriguing to combine novel platforms, such as twisted van der Waals materials, with the field of cavity engineering. On the subject of superradiance we expect nonequilibrium superradiance to move into the center of attention. Finally, the investigation of entangled photons in cavities as a probe of the quantum material under scrutiny or in their own right for applications in quantum technologies could prove fruitful.  
}
\label{fig.outlook}
\end{figure*}

\section{Outlook}
\label{sec:outlook}
Cavity quantum materials is clearly an emergent research field in its very infancy. It brings together researchers from a variety of communities, including quantum materials science (especially in two-dimensions), quantum many-body physics, ultrafast and nonlinear laser-based spectroscopy, quantum optics and semiconductor physics, nanoplasmonics and nanophotonics, and polaritonic chemistry. Naturally, this infancy implies that community-building measures are of tremendous importance to find a common language and identify relevant near- and midterm goals. In the following we sketch our personal vision for the field, summarized in Fig.~\ref{fig.outlook}.

One key aspect that has become evident in this brief review is that it is mostly theory-driven, with only few subfields where experiments are charting the path. Therefore theoretical proposals with clear predictions and guidelines for experimentalists are needed. In order to achieve this, theory needs to become more realistic in treating the light field: descriptions need to go beyond the long-wavelength limit, include multimode descriptions, take the proper cavity geometry into account, involve near-field effects and a realistic inclusion of screening and light-matter feedback effects. To this end, it is desirable to join forces with \textit{ab initio} descriptions of light and matter in polaritonic chemistry that are being developed \cite{sidler_perspective_2021} in order to achieve predictive power for the control of quantum materials with classical and quantum light \cite{lloyd-hughes_2021_2021}.

Regarding the control of materials properties via collective modes, it is important to devise strategies to employ cavities to induce mode softening and strong fluctuations outside the light cone, that is, at wave vectors that span a sizeable fraction of the Brillouin zone. Such strategies naturally connect to the above-mentioned finite-wavelength and near-field effects and would allow to trigger phase transitions, e.g., cavity-induced ferroelectricity, charge-density wave formation, or even superradiance. As discussed above, superradiance-like phenomena could also potentially be generated by driving a cavity-material platform away from thermal equilibrium, and thus the demonstration of non-equilibrium superradiance in quantum materials inside a cavity constitutes one important goal in the field.  

More generally, we envision that there is ``plenty of room'' in between the quantum limit of dark cavities at strong light-matter coupling and the classical limit of strongly driven systems. As described above, first theoretical results indicate that the vision of Floquet engineering of many-body properties in quantum materials might be achievable in a cavity, in which heating and decoherence issues could be mitigated. Therefore we posit that one central goal of the field is to explore the crossover regime between quantum and classical Floquet engineering and to demonstrate few-photon Floquet engineering. 

From the quantum materials perspective, it is desirable to synthesize new materials that engineer favorable light-matter coupling effects. As briefly mentioned above, Moir\'e van der Waals materials are one promising class of systems for this endeavor, as they allow for tunability of electronic properties and have shown intriguing potential to realize various prototypical solid state model Hamiltonians \cite{kennes_moire_2021}. Their controllable and nontrivial quantum-geometric multi-band properties can lead to outsized light-matter coupling effects in quasi-flat electronic bands with squeezed kinetic energies \cite{PhysRevB.104.064306}. This would establish {\it cavitwistronics} as the combinations of twistronics with cavitronics.

Moreover, cavity quantum materials could be used as photonic platforms and integrated in photon-based quantum technologies. 
We envision that the strong electronic interactions which are typical of quantum materials could provide a means to create efficient photonic interactions for two-photon quantum gates, and enable the creation of nonclassical states of light.  With the notable exception of few experiments in Landau polaritons~\cite{Mornhinweg2021} or molecular samples~\cite{Wang2021}, this direction of research has remained largely unexplored to date. 

Finally, it is an intriguing perspective to employ the strong coupling between photons and matter as a sensor, or probe, for the emergent many-body properties of the quantum materials embedded in cavities. This could for instance help identify fluctuations near phase transitions and -- due to the collective nature of the cavity-matter coupling -- provide access to nonlocal information about the system, such as entanglement entropies or topological properties. For instance, recent analytical work and matrix product state simulations connect the R\'enyi entropy to the cavity Fano factor in the case of a Kitaev chain coupled to a single cavity mode \cite{MendezCordoba2020}. 
It remains to be seen whether these promising insights can be transferred to the analysis of other correlated materials and establish cavity-enabled fluctuation sensing as a novel probe in condensed matter platforms.

\begin{acknowledgments}
Discussions with and input by Silvia Viola Kusminskiy are gratefully acknowledged. We thank Christian Eckhardt for proofreading and valuable feedback, and Francesco Piazza for helpful comments.
FS acknowledges support from the Cluster of Excellence 'Advanced Imaging of Matter' of the Deutsche Forschungsgemeinschaft (DFG) - EXC 2056 - project ID 390715994. DMK acknowledges the Deutsche Forschungsgemeinschaft (DFG, German Research Foundation) for support through RTG 1995 and under Germany's Excellence Strategy - Cluster of Excellence Matter and Light for Quantum Computing (ML4Q) EXC 2004/1 - 390534769.
MAS acknowledges financial support through the Deutsche Forschungsgemeinschaft (DFG, German Research Foundation) via the Emmy Noether program (SE 2558/2). DMK and MAS acknowledge support from the Max Planck-New York City Center for Non-Equilibrium Quantum Phenomena.
\end{acknowledgments}

%\nocite{*}
\bibliography{000_cavity_review,Cavity-Review,magnonics,newbib}% Produces the bibliography via BibTeX.

\end{document}